\documentclass{main}
\usepackage{amsmath}
\usepackage{amssymb}

\def\be{\begin{equation}}
\def\ee{\end{equation}}
\def\bea{\begin{eqnarray}}
\def\eea{\end{eqnarray}}

\newcommand{\Pomeron}{I\!\!P}

\newcommand{\Photo}{\includegraphics[height=35mm]{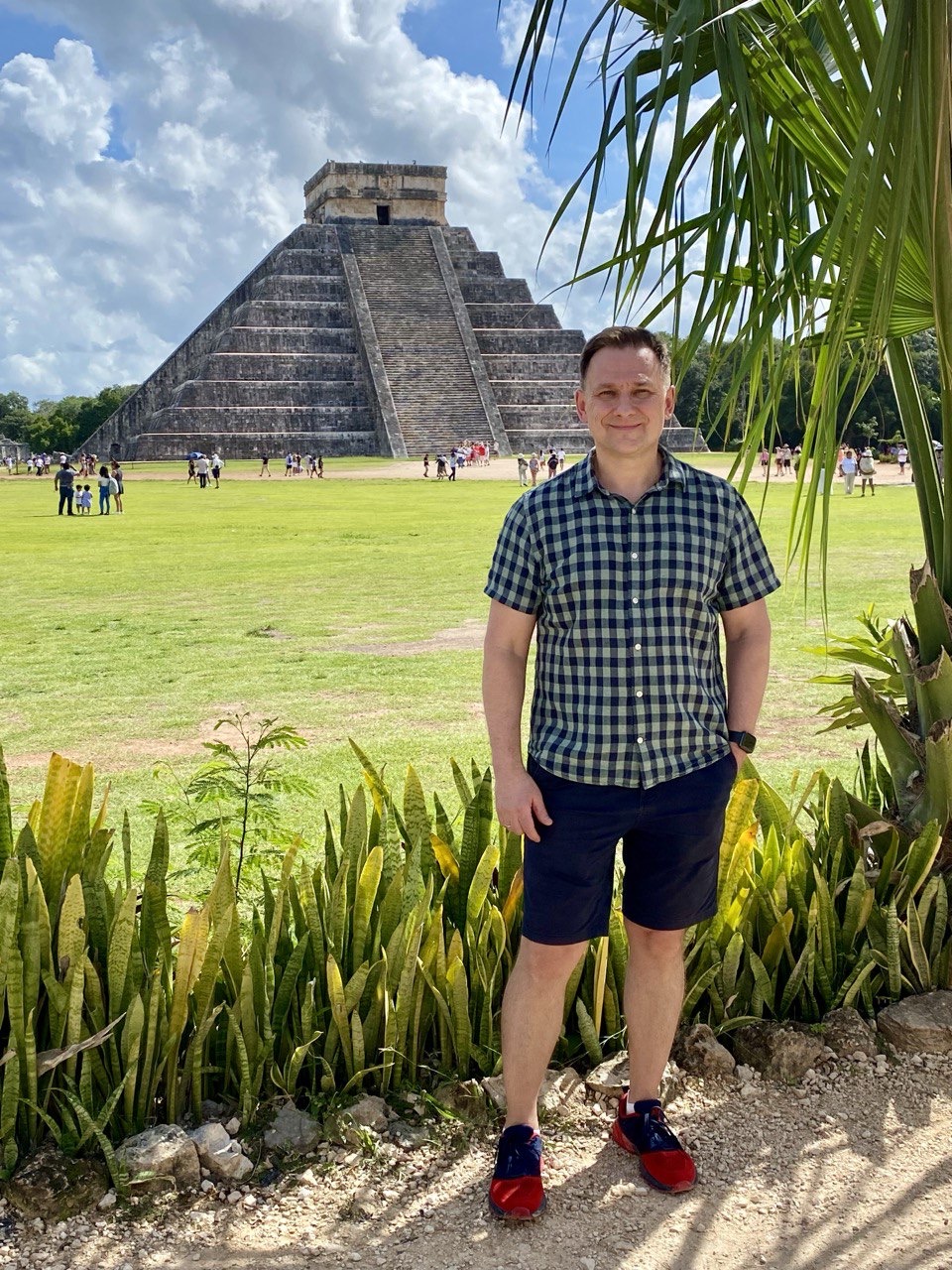}}

\begin{document}

\title{\Large INCLUSIVE AND DIFFRACTIVE DIJET PHOTOPRODUCTION IN ULTRAPERIPHERAL Pb-Pb COLLISIONS AT THE LHC}

\author{V. GUZEY}

\address{University of Jyvaskyla, Department of Physics, P.O. Box 35, FI-40014 University
of Jyvaskyla, Finland and Helsinki Institute of Physics, P.O. Box 64, FI-00014 University of Helsinki, Finland}

\maketitle\abstracts{
In this contribution, we summarize NLO pQCD predictions for inclusive and diffractive dijet photoproduction in Pb-Pb UPCs at the LHC. 
We demonstrate that the theory describes well the preliminary ATLAS data on the inclusive cross section, which probes nuclear parton distributions (PDFs) down
to $x_A \approx 0.005$ and which can reduce current uncertainties of the small-$x$ nuclear gluon distribution by approximately a factor of 2.
Employing predictions of the leading twist approach to nuclear shadowing for nuclear diffractive PDFs, we calculate the cross section of 
diffractive dijet photoproduction and show that its $x_{\gamma}$ dependence is sensitive to the effect of nuclear shadowing and the mechanism of QCD factorization breaking in hard diffraction.
We also find that due to large leading twist nuclear shadowing and restricted kinematics, the diffractive contribution to the inclusive cross section of dijet photoproduction does not exceed $5-10$\%, which helps with an ambiguous interpretation of the ATLAS data.
}
  

\section{Introduction}
\label{sec:intro}

Jets are collimated sprays of hadrons ($\pi$, $K$, $\dots$) produced in high-energy $e^{+} e^{-}$, lepton-hadron, and 
hadron-hadron collisions. Jets have been instrumental in establishing quantum 
chromodynamics (QCD) as the correct theory of the strong interactions and, in particular, its concepts of
asymptotic freedom and confinement~\cite{Salam:2010zt,Sapeta:2015gee,Laenen:2016njs}. A classic example of it is 3-jet events observed in $e^{+} e^{-}$ annihilation, which has proved the existence of gluons. In perturbative QCD (pQCD), measurements of jets are commonly used to test validity of the QCD collinear factorization theorem,
determine the strong coupling constant $\alpha_s$~\cite{ZEUS:2012pcn,H1:2014cbm}, and provide complementary information on parton distribution functions (PDFs)~\cite{Klein:2008di,Newman:2013ada}.

In particular, since in electron-proton ($ep$) deep-inelastic scattering (DIS) and hard hadron scattering, jet cross sections are sensitive
to quark and gluon distributions of the target at the same order of the perturbation series in powers of $\alpha_s$, jet data provide additional constraints on the gluon distribution, which complement those from the total DIS cross section.  
Global QCD fits of proton PDFs take advantage of it by including data on jet production in $ep$ DIS at HERA, proton-antiproton 
scattering at Tevatron and proton-proton ($pp$) scattering at the Large Hadron Collider (LHC)~\cite{ZEUS:2005iex,NNPDF:2021njg,Hou:2019efy}.
Extending this to nuclear targets, essential constraints on the nuclear gluon distribution have been obtained by employing data on dijet production in proton-nucleus scattering 
at the LHC~\cite{Eskola:2021nhw}. Similarly, HERA data on dijet photoproduction on the proton enables one to determine the gluon
distribution in the real photon more reliably compared to the case, when one only uses the data on the $F_{2}^{\gamma}(x,Q^2)$ 
structure function measured in $e^{+} e^{-}$ annihilation~\cite{Slominski:2005bw}.

In addition, it has been discussed in the literature that measurements of forward dijet production may aid in 
searching for small-$x$ Balitsky-Fadin-Kuraev-Lipatov (BFKL) and saturation physics at the LHC~\cite{Hentschinski:2022xnd,Iancu:2023lel} and the planned Electron-Ion Collider (EIC) in USA~\cite{Boussarie:2021ybe}.
Finally, jets present a Standard Model background for many new physics processes.

Focusing on production of jets by real photons, we notice that while all experimental information on jet photoproduction comes from 
$ep$ scattering at HERA~\cite{Klein:2008di,Newman:2013ada,Butterworth:2005aq}, there is preliminary ATLAS data on dijet photoproduction 
in Pb-Pb ultraperipheral collisions (UPCs) at the LHC~\cite{ATLAS:2017kwa,ATLAS:2022cbd}. In UPCs, colliding ions pass each other at large impact parameters and interact via emission of quasi-real photons, which effectively makes the LHC a high-energy and high-intensity photon-nucleus 
collider~\cite{Baltz:2007kq}. So far the emphasis of UPC measurements has been coherent and incoherent production of light and heavy vector mesons.
Notably, it has been argued that exclusive photoproduction of charmonium $J/\psi$ mesons in $pp$ UPCs  allows one to probe and constrain the gluon density in the proton down to $x_p \approx 3 \times 10^{-6}$ at the resolution scale of the order of the charm quark mass~\cite{Flett:2020duk}.
In the nucleus case, the data on coherent $J/\psi$ photoproduction in Pb-Pb UPCs have discovered a large nuclear suppression, which
can be interpreted in terms of strong gluon nuclear shadowing~\cite{Guzey:2013xba,Guzey:2013qza} at $x_A \approx 10^{-3}$ and all the way down to $x_A \approx 10^{-5}$. These findings have nicely confirmed predictions of the leading twist approach (LTA) to nuclear shadowing~\cite{Frankfurt:2011cs}.

Measurements of inclusive and diffractive dijet photoproduction in Pb-Pb UPCs at the LHC allow one to expand the scope of the UPC physics program.
Compared to $J/\psi$ production, the theoretical description of the dijet cross section in perturbative QCD is cleaner since it does not involve the charmonium wave function and complications associated with the kinematics of exclusive reactions and generalized parton distributions.
In the case of inclusive dijet production, $A+A \to A + 2{\rm jets}+X$, where $X$ denotes the hadronic final state resulting from nucleus dissociation,
the dijet cross section probes nuclear and real photon PDFs at large energy scales, which are determined by the jet 
transverse momentum and, thus, explores the kinematic region complementary to the one in vector meson photoproduction.
Requiring that the target nucleus is intact, one can study diffractive dijet photoproduction, $A+A \to A + 2{\rm jets}+X^{\prime}+A$, which gives an access to novel nuclear diffractive PDFs
and which may also shed new light on the mechanism of QCD factorization breaking in hard diffraction. 

Pioneering leading-order (LO) pQCD calculations of photoproduction of heavy flavor (bottom) jets~\cite{Strikman:2005yv} and heavy quarks~\cite{Klein:2002wm} (see also~\cite{Goncalves:2009ey,Goncalves:2017zdx}) in UPCs have demonstrated feasibility and large rates of such measurements in the LHC kinematics..

The rest of this contribution is organized as follows. Section~\ref{sec:dijet_inclusive} summarizes NLO pQCD predictions 
for the cross section of inclusive dijet photoproduction in Pb-Pb UPCs at 5.02 TeV, their comparison to the ATLAS data and the magnitude of nuclear modifications as well as the potential
of this process to provide additional constraints on nuclear PDFs at small $x$. In Sec.~\ref{sec:dijet_diffractive}, we present NLO pQCD predictions for diffractive
dijet photoproduction in Pb-Pb UPCs at the LHC and discuss sensitivity of the $x_{\gamma}$ dependence to the mechanism of the QCD factorization 
breaking in hard diffraction. We also quantify the diffractive contribution to the cross section of inclusive dijet photoproduction in UPCs.
We draw conclusions in Sec.~\ref{sec:conclusions}.

\section{Inclusive dijet photoproduction in Pb-Pb UPCs at the LHC}
\label{sec:dijet_inclusive}

\begin{figure}[t]
  \centerline{%
    \includegraphics[width=12cm]{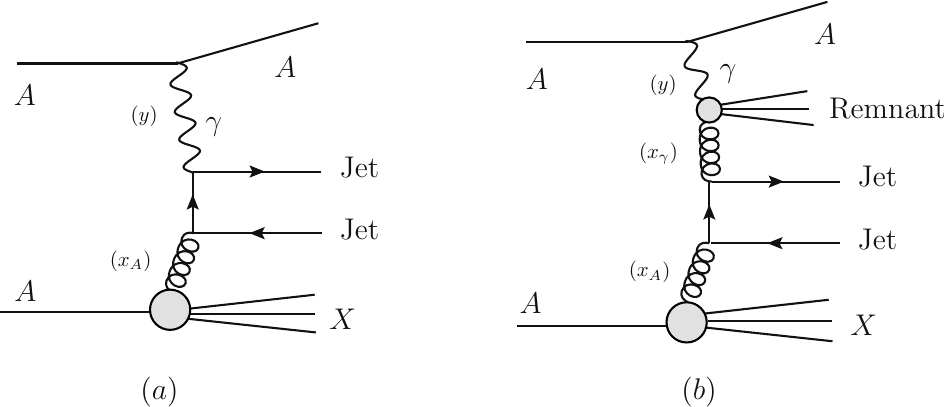}}
  \caption{Typical LO pQCD diagrams for inclusive dijet photoproduction in $AA$ UPCs: (a) direct and (b) resolved photon contributions. 
  The corresponding momentum fractions are given in parenthesis.
  }
\label{fig:dijets_upc}
\end{figure}

In the framework of collinear factorization of pQCD, the formalism for calculation of the cross section of inclusive dijet photoproduction is well-established, and results of calculations performed at next-to-leading order (NLO) accuracy successfully describe the available HERA data~\cite{Frixione:1997ks,Klasen:1996it,Aurenche:2000nc,Klasen:2002xb}. Applying it to UPCs, one can write the cross section of inclusive dijet photoproduction in Pb-Pb UPCs, $A+A \to A + 2{\rm jets}+X$, as the following convolution~\cite{Guzey:2018dlm}, 
\begin{equation}
d\sigma(AA \to A + 2{\rm jets} +X) = \sum_{a,b} \int dy \int dx_{\gamma} \int dx_A f_{\gamma/A}(y) f_{a/\gamma}(x_{\gamma},\mu) f_{b/A}(x_A,\mu) d\hat{\sigma}_{ab \to \rm jets} \,,
\label{eq:dsigma}
\end{equation}
where $f_{\gamma/A}(y)$ is the flux of equivalent photons, $f_{a/\gamma}(x_{\gamma},\mu)$ are photon PDFs, $f_{b/A}(x_A,\mu)$ are nuclear PDFs,
and $d\hat{\sigma}_{ab \to \rm jets}$ is the elementary cross section to produce jets in hard scattering of partons $a$ and $b$.
The longitudinal momentum fractions are $y$ for the photon, $x_{\gamma}$ for parton $a$ in the photon, and $x_A$ for parton $b$ in a target nucleus.
The PDFs are evaluated at the resolution scale $\mu$, which is usually identified with the jet transverse momentum $p_T$.

Figure~\ref{fig:dijets_upc} illustrates Eq.~(\ref{eq:dsigma}) by showing typical LO pQCD diagrams for the direct contribution (graph $a$), where the photon as a whole takes part in the hard reaction, and the resolved contribution (graph $b$), where a parton of a hadronic fluctuation of the photon
(photon PDFs) participates in the hard scattering reaction. The corresponding momentum fractions are given in parenthesis.
Separation of the direct and resolved contributions is unambiguous only at LO, where
$f_{a/\gamma}(x_{\gamma},\mu)=\delta(1-x_{\gamma})$ for the direct term. At NLO, due to renormalization of standard collinear divergences of pQCD, 
the definition of the direct and resolved terms begins to depend on a choice of the factorization scheme and scale. Nevertheless, the notion of the direct and resolved contributions remains useful.  

The photon flux $N_{\gamma/A}(y)$ is usually calculated using the Weizs\"acker-Williams equivalent photon approximation combined with the probability for the nuclei not to interact strongly at small impact parameters. However, for purposes of the present analysis, the exact  
expression can be approximated by the photon flux produced by a relativistic point-like charge $Z$ passing a target at the minimum impact parameter 
$b_{\rm min}$,
\begin{equation}
N_{\gamma/A}(y)=\frac{2 \alpha_{\rm e.m.}Z^2}{\pi} \frac{1}{y} \left[\zeta K_0(\zeta)K_1(\zeta)-\frac{\zeta^2}{2} (K_0^2(\zeta)-K_1^2(\zeta))\right] \,,
\label{eq:flux}
\end{equation}
where $\alpha_{\rm e.m.}$ is the fine-structure constant, $K_{0,1}$ are modified Bessel functions of the second kind, 
and $\zeta=y m_p b_{\rm min}$ with $m_p$ the proton mass and $b_{\rm min}=14.2$ fm for Pb-Pb UPCs~\cite{Nystrand:2004vn}.

\begin{figure}[t]
  \centerline{%
    \includegraphics[width=8cm]{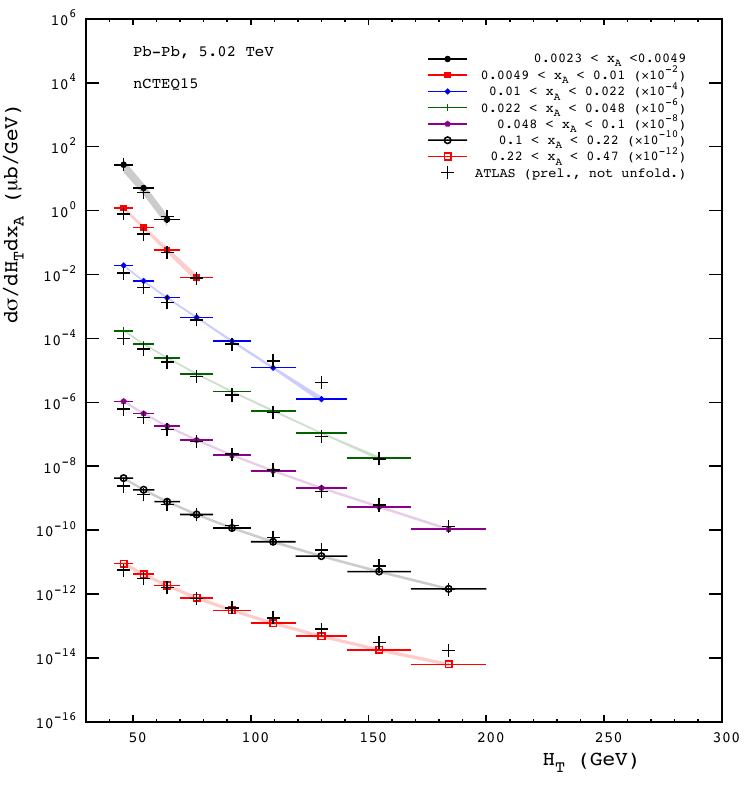}
    \includegraphics[width=8cm]{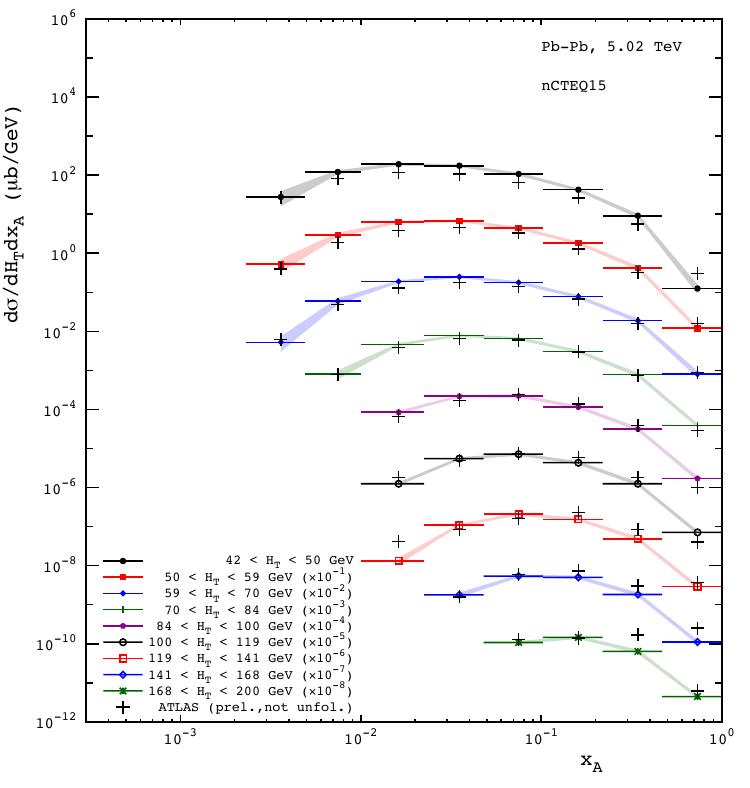}}
  \caption{The cross section of inclusive dijet photoproduction in Pb-Pb UPCs at $\sqrt{s_{NN}}=5.02$ TeV as a function of $H_T=p_{T,1}+p_{T,2}$ in different bins of 
  $x_A$ (left) and as a function of $x_A$ in bins of $H_T$ (right). The NLO pQCD predictions with the nCTEQ15 nuclear PDFs are compared with the ATLAS preliminary data, see text for details.
}
  \label{fig:NLO_vs_ATLAS}
\end{figure}

The momentum fractions $x_{\gamma}$ and $x_A$ in Eq.~(\ref{eq:dsigma}) can be estimated using the measured jet rapidities and transverse momenta.
In particular, one can use the following relations
\begin{equation}
x_{\gamma}=\frac{m_{\rm jets}}{\sqrt{s_{NN}}} e^{y_{\rm jets}} \,, \quad \quad
x_{A}=\frac{m_{\rm jets}}{\sqrt{s_{NN}}} e^{-y_{\rm jets}} \,,
\label{eq:x_A}
\end{equation}
where $m_{\rm jets}$ is the invariant mass of the jet system and $y_{\rm jets}$ its rapidity~\cite{ATLAS:2017kwa}.

Note that Eq.~(\ref{eq:dsigma}) gives the cross section at the level of massless partons and for a comparison with data needs to be supplemented with
hadronization corrections. They are usually estimated using Monte Carlo generators, which involve LO matrix elements and include the effects of 
parton showers~\cite{Helenius:2018mhx,Helenius:2019gbd}.

Figure~\ref{fig:NLO_vs_ATLAS} presents a comparison of the NLO pQCD predictions~\cite{Guzey:2018dlm} for the cross section of inclusive dijet photoproduction
in Pb-Pb UPCs at $\sqrt{s_{NN}}=5.02$ TeV with the preliminary ATLAS data~\cite{ATLAS:2017kwa}. The calculation used the nCTEQ15 nuclear PDFs~\cite{Kovarik:2015cma}, whose uncertainty propagation is shown by the shaded bands, GRVHO photon PDFs~\cite{Gluck:1991jc}, and
the ATLAS experimental cuts, 
importantly, $p_{T,1} > 20$ GeV for the leading jet and $p_{T,2} > 15$ GeV for other jets.
The left panel shows the distribution in
$H_T=p_{T,1}+p_{T,2}$ in different bins of $x_A$; the right panel shows the distribution
in $x_A$ in bins of $H_T$. One can see from the figure that the NLO pQCD results correctly reproduce the shape and normalization
of the data, which have not been corrected for detector response. Note that in the considered kinematics, the coverage in $x_A$ extends down to 
$x_A \approx 5 \times 10^{-3}$ at $\mu^2=(H_T/2)^2$, which provides a certain sensitivity to nuclear PDFs at small $x$ and large $\mu$, see the discussion below.

\begin{figure}[t]
  \centerline{%
    \includegraphics[width=8cm]{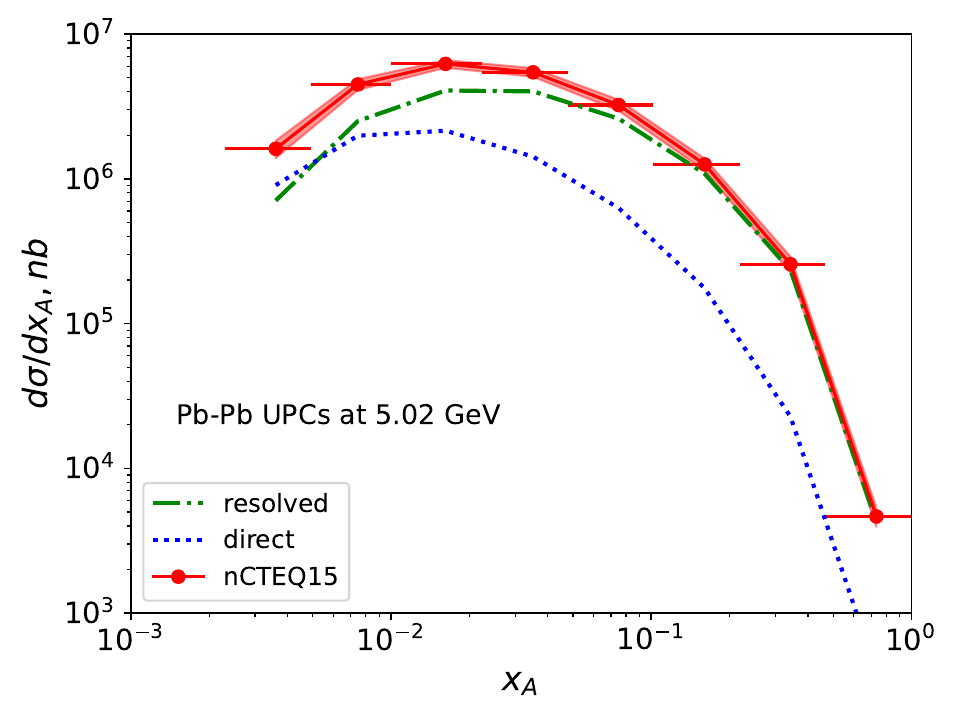}
    \includegraphics[width=8cm]{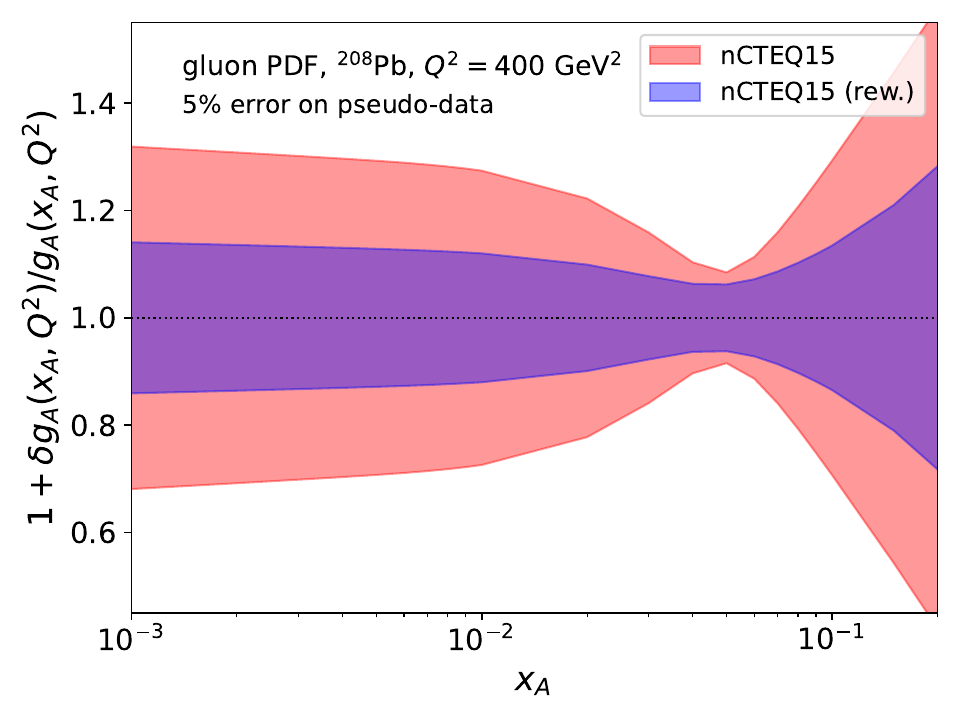}}
  \caption{(Left) NLO pQCD predictions for the cross section of dijet photoproduction in Pb-Pb UPCs at 5.02 TeV as a function of $x_A$ 
  using the nCTEQ15 nuclear PDFs and ATLAS experimental cuts. The curves show separately the direct (dashed) and resolved (dot-dashed) photon
  contributions as well as their sum (solid). (Right) Relative uncertainties of the nCTEQ15 gluon distribution in $^{208}$Pb as a function of $x_A$ at
  $Q^2=400$ GeV$^2$ before (outer pink shaded band) and after
(inner blue band) the reweighting using pseudo-data on dijet photoproduction in Pb-Pb UPCs at 5.02 TeV.
}
  \label{fig:rew}
\end{figure}

It is instructive to examine an interplay of the direct and resolved photon contributions to the dijet cross section. 
The left panel of Fig.~\ref{fig:rew} shows separately the direct and resolved terms as well as their sum as a function of $x_A$. One can see from the figure that the resolved contribution dominates for $x_A > 0.01$, while the two contributions are compatible in size for $x_A < 0.01$.
This trend is generally expected because the direct photon contribution increases in the $x_{\gamma} \to 1$ limit (the values of 
$x_{\gamma}$ and $x_A$ are anti-correlated, see Eq.~(\ref{eq:x_A})) and agrees with the expectations based on PYTHIA 8 Monte Carlo framework~\cite{Helenius:2018mhx,Helenius:2019gbd}.

The cross section of dijet photoproduction in Pb-Pb UPCs in Fig.~\ref{fig:rew} is sensitive to nuclear modifications of nuclear PDFs: 
its $\approx 10$\% suppression for $x_A < 0.01$ compared to the impulse approximation estimate is caused by nuclear shadowing and $\approx 20$\% enhancement around $x_A < 0.1$ by antishadowing of the gluon distribution; it is followed by a $5-10$\% suppression for $x_A > 0.05$ due to
the EMC effect, which is built in most of nuclear PDFs.

One can turn this around and quantitatively investigate the potential of this process to further constrain nuclear PDFs using the technique of
Bayesian reweighting~\cite{Guzey:2019kik}. In short, using $N_{\rm rep}=10,000$ random replicas of nCTEQ15 error PDFs, $f^{k}_{i/A}(x,Q^2)$ with $i$ the parton flavor, one can calculate the dijet cross section for each replica $k=1, 2, \dots, N_{\rm rep}$ and then determine the statistical weights $\omega_k$ for replicas to reproduce observables.
The latter was taken to be the dijet cross section calculated using the central nCTEQ15 PDFs, which was assigned a $5-15$\% uncertainty playing
the role of experimental errors. After this procedure, the new reweighted central values $\langle f_{i/A}(x,Q^2)\rangle_{\rm new}$ and 
uncertainties $\delta \langle f_{i/A}(x,Q^2)\rangle_{\rm new}$ of nuclear PDFs become
\begin{eqnarray}
\langle f_{i/A}(x,Q^2)\rangle_{\rm new} &=&\frac{1}{N_{\rm rep}} \sum_{k=1}^{N_{\rm rep}} \omega_k f^{k}_{i/A}(x,Q^2) \,, \nonumber\\
\delta \langle f_{i/A}(x,Q^2)\rangle_{\rm new} &=&\sqrt{\frac{1}{N_{\rm rep}} \sum_{k=1}^{N_{\rm rep}} \omega_k 
\left(f^{k}_{i/A}(x,Q^2)-\langle f_{i/A}(x,Q^2)\rangle_{\rm new}\right)^2} \,.
\end{eqnarray}

The right panel of Fig.~\ref{fig:rew} illustrates the result of the Bayesian reweighting of the nCTEQ15 gluon distribution in $^{208}$Pb at 
$Q^2=400$ GeV$^2$ using the procedure outlined above. It presents 
the relative uncertainty of the gluon nPDF, $1+\delta g_{A}(x_A,Q^2)/g_{A}(x_A,Q^2)$, as a function of $x_A$ before (outer pink shaded band) and after
(inner blue band) the reweighting. One can see from the 
figure that assigning a 5\% uncertainty to the pseudo-data leads to a reduction of uncertainties in the gluon distribution for $x_A < 0.005$ by approximately a factor of 2.

\section{Diffractive dijet photoproduction in Pb-Pb UPCs at the LHC}
\label{sec:dijet_diffractive}

\begin{figure}[t]
  \centerline{%
    \includegraphics[width=12cm]{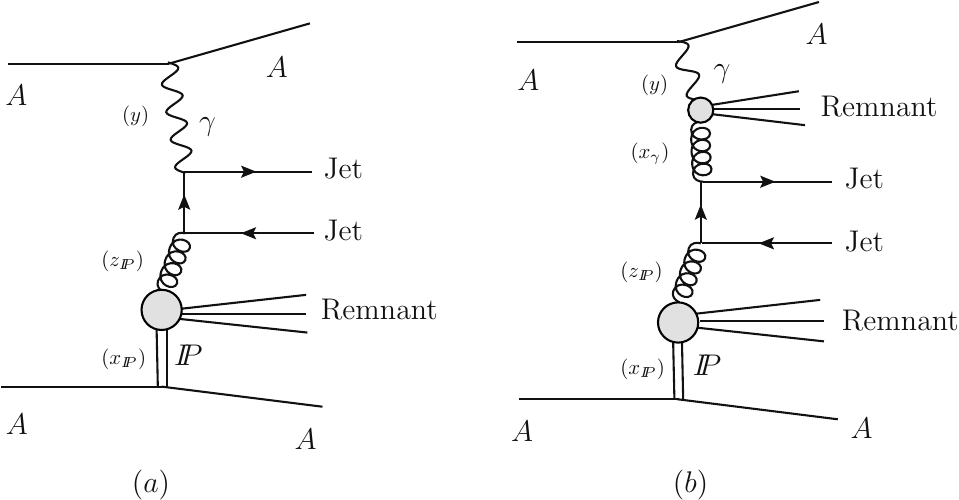}}
  \caption{Typical LO pQCD diagrams for diffractive dijet photoproduction in $AA$ UPCs: (a) direct and (b) resolved photon contributions. 
  The diffractive exchange is denoted by the Pomeron contribution.
  The corresponding momentum fractions are given in parenthesis.
  }
\label{fig:dijets_upc_diff}
\end{figure}

The considerations of the previous section can be extended to diffractive jet photoproduction in heavy-ion UPCs, where one imposes the additional condition that the target nucleus remains intact and recoils elastically. Figure~\ref{fig:dijets_upc_diff} shows typical LO pQCD diagrams for the direct
(graph $a$) and resolved (graph $b$) photon contributions, where the 
vertical double lines denote the diffractive exchange (Pomeron flux) labeled ``$\Pomeron$''.

Unlike the inclusive case, UPC cross sections corresponding to coherent underlying photon-nucleus scattering receive contributions from both 
left-moving and right-moving colliding ions, which introduces a well-known two-fold ambiguity between the photon energy and rapidity of the final dijet system.
Generalizing Eq.~(\ref{eq:dsigma}), the contribution of the right-moving photon source to the cross section of diffractive dijet photoproduction
in Pb-Pb UPCs, $A+A \to A + 2{\rm jets}+X^{\prime}+A$, where $X^{\prime}$ includes hadronic debris of the ``Pomeron'' and photon,
 can be written in the following form~\cite{Guzey:2016tek}
\begin{eqnarray}
d\sigma(AA \to A + 2{\rm jets} +X^{\prime}+A)^{(+)} &=& \sum_{a,b} \int dt \int dx_{\Pomeron} \int dz_{\Pomeron} \int dy \int dx_{\gamma} \nonumber\\
&\times&
 f_{\gamma/A}(y) f_{a/\gamma}(x_{\gamma},\mu) f_{b/A}^{D(4)}(x_{\Pomeron},z_{\Pomeron},t,\mu) d\hat{\sigma}_{ab \to \rm jets} \,.
\label{eq:dsigma_diff}
\end{eqnarray}
The contribution of the left-moving photon source is obtained from Eq.~(\ref{eq:dsigma_diff}) by inverting signs of the jet rapidities. 
In Eq.~(\ref{eq:dsigma_diff}), $f_{b/A}^{D(4)}(x_{\Pomeron},z_{\Pomeron},t,\mu)$ denote the nuclear diffractive PDFs, which represent the conditional probability to find parton $b$ with the momentum fraction $z_{\Pomeron}$ with respect to the diffractive exchange (Pomeron) carrying the momentum
fraction $x_{\Pomeron}$, provided that the nucleus remains intact and recoils elastically with the momentum transfer $t$.
The momentum fractions involved in Eq.~(\ref{eq:dsigma_diff}) are shown in parenthesis in Fig.~\ref{fig:dijets_upc_diff}.

Similarly to usual nuclear PDFs, nuclear diffractive PDFs are subject to nuclear modifications, notably, due to nuclear shadowing.
The leading twist approach to nuclear shadowing~\cite{Frankfurt:2011cs} makes definite predictions for $f_{b/A}^{D(4)}$, which are characterized by their strong suppression at small $x$. It can be quantified by introducing the suppression factor $R_b$ with respect to $f_{b/A}^{D(4)}$ evaluated in the impulse approximation,
\begin{equation}
f_{b/A}^{D(4)}(x_{\Pomeron},z_{\Pomeron},t,\mu)=R_b(x_{\Pomeron},z_{\Pomeron},\mu) A^2 F_A^2(t) f_{b/p}^{D(4)}(x_{\Pomeron},z_{\Pomeron},t=0,\mu) \,,
\label{eq:R_b}
\end{equation}
where $F_A(t)$ is the nucleus form factor and $f_{b/p}^{D(4)}$ is the diffractive PDFs of the proton. An examination of $R_b$ shows that it rather
weakly depends on flavor $b$, the light-cone momentum fractions $x_{\Pomeron}$ and $z_{\Pomeron}$ (provided that $x_{\Pomeron}$ is sufficiently small), 
and the resolution scale $\mu$. Hence, for the purpose of estimating yields of this process, one can approximate $R_b$ by a single number~\cite{Guzey:2016tek,Guzey:2024xpa}
\begin{equation}
R_b(x_{\Pomeron},z_{\Pomeron},\mu) \approx 0.08-0.16 \,,
\label{eq:R_b2}
\end{equation}
where the upper and lower values correspond to the ``high shadowing'' and ``low shadowing'' scenarios, respectively. This
spread in the values for $R_b$ reflects a significant theoretical uncertainty of LTA predictions for nuclear diffractive PDFs, for a detailed discussion, see~\cite{Guzey:2024xpa}. 

\begin{figure}[t]
  \centerline{%
    \includegraphics[width=8cm]{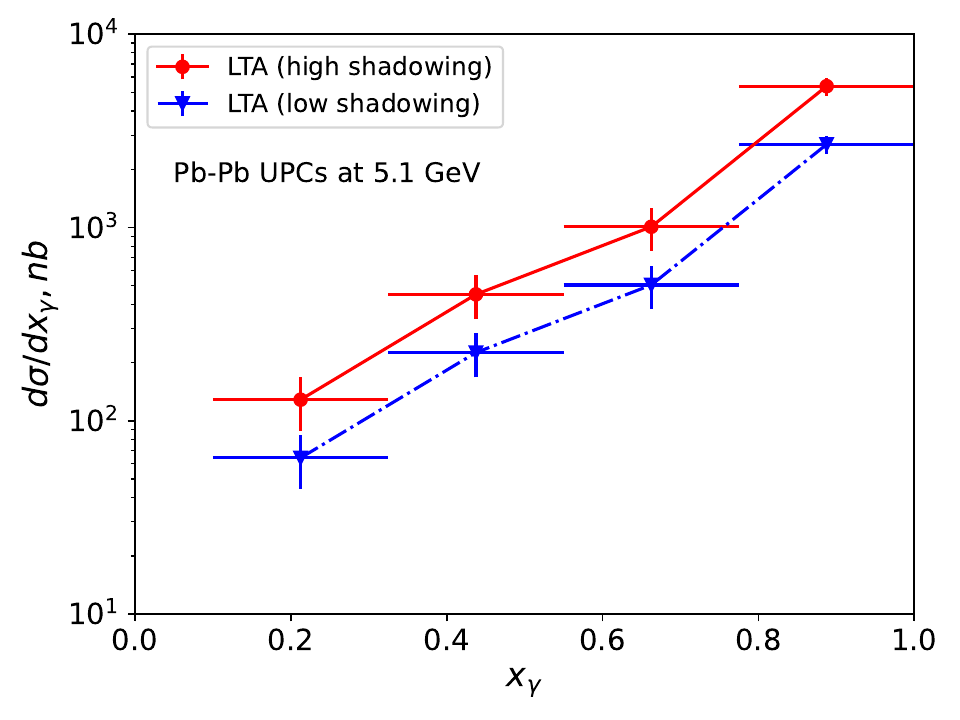}
    \includegraphics[width=8cm]{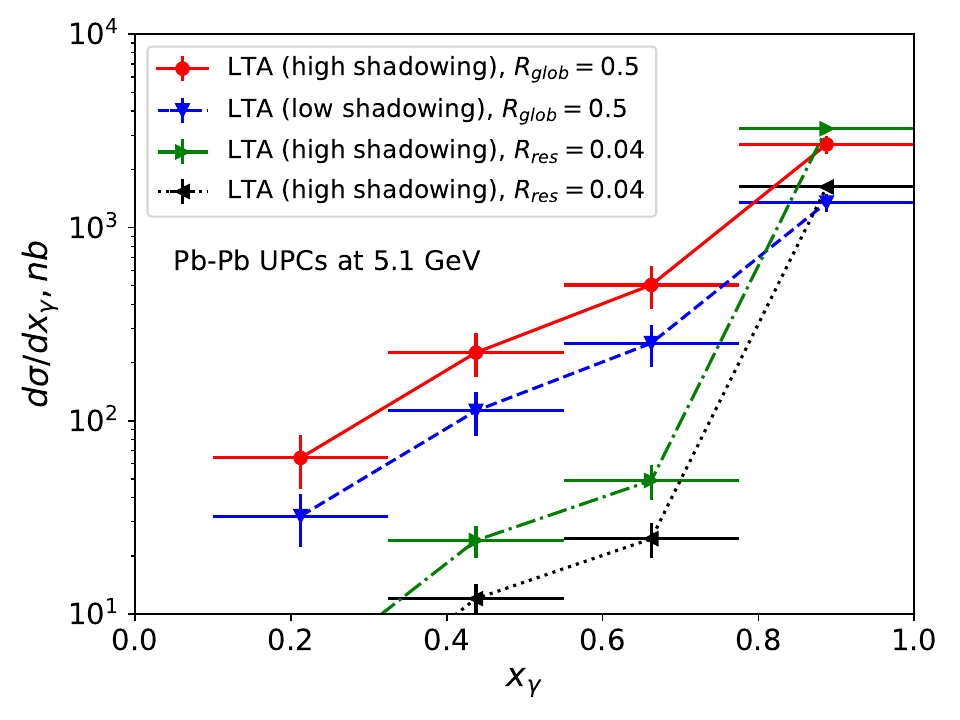}}
  \caption{NLO pQCD predictions for the cross section of diffractive dijet photoproduction in Pb-Pb UPCs at 5.1 TeV as a function of $x_{\gamma}$.
   The results are obtained using LTA predictions for nuclear diffractive PDFs (``high shadowing'' and ``low shadowing'' versions) and
   assuming either no QCD factorization breaking in hard diffraction (left) or two scenarios of it realized through the suppression factors $R_{\rm glob}$ and 
   $R_{\rm res}$ (right), see text for details.
}
  \label{fig:dsigma_diff}
\end{figure}

The left panel of Fig.~\ref{fig:dsigma_diff} presents NLO pQCD predictions for the cross section of diffractive dijet photoproduction in Pb-Pb UPCs at 5.1 TeV as a function of $x_{\gamma}$ using the LTA results for nuclear diffractive PDFs, see Eqs.~(\ref{eq:R_b}) and (\ref{eq:R_b2}), 
$p_{T,1} > 20$ GeV for the leading jet and $p_{T,2} > 18$ GeV for sub-leading jets and other otherwise 
generic cuts specified in~\cite{Guzey:2016tek}. 
The upper and lower curves corresponds to the ``high shadowing'' and ``low shadowing'' LTA predictions, respectively.
The horizontal lines show the width of bins in $x_{\gamma}$, and the vertical bars quantify the effect of the variation of the hard scale of the process in the 
$p_{T,1}/2 < \mu < 2 p_{T,1}$ interval. One can see from the figure that the uncertainty due to the scale variation is smaller than the uncertainty in the value of $R_b$. Overall, the figure shows that the predicted yields are significant demonstrating feasibility of such measurements.

Analyses of diffractive dijet photoproduction in $ep$ scattering at HERA have shown that the QCD factorization theorem for hard diffraction~\cite{Collins:1997sr} is violated and NLO pQCD calculations overestimate the cross sections measured by the ZEUS and H1 collaborations at HERA  by a approximately a factor of 2, see details in~\cite{Klasen:2010vk}. The pattern of this factorization breaking is yet unknown since a good description of the data can be achieved by introducing
a global suppression factor $R_{\rm glob}=0.5$ or the suppression factor $R_{\rm res}=0.34$ for the resolved photon contribution only, or the flavor-dependent 
and $x_{\gamma}$-dependent suppression factor interpolating between $R_{\rm glob}$ and $R_{\rm res}$, see~\cite{Guzey:2016awf}.

As a consequence, one of the most sensitive observables to various scenarios of factorization breaking is the distribution in the momentum 
fraction $x_{\gamma}$. Thus, it has been argued in~\cite{Guzey:2016tek} that measurements of the $x_{\gamma}$ dependence of the cross section of 
diffractive dijet photoproduction in UPCs at the LHC may help to shed new light on this phenomenon.

The right panel of Fig.~\ref{fig:dsigma_diff} illustrates the effect of the QCD factorization breaking for hard diffraction on NLO pQCD predictions for 
the cross section of diffractive dijet photoproduction in Pb-Pb UPCs at 5.1 TeV as a function of $x_{\gamma}$. These predictions are obtained by rescaling
the result of Eq.~(\ref{eq:dsigma_diff}) either by the global suppression factor of $R_{\rm glob}=0.5$ or by suppressing only the resolved photon term by 
the factor of $R_{\rm res}=0.04$. The latter value is estimated using the Glauber model for $\rho$ meson-nucleus scattering, see details in~\cite{Guzey:2016awf}.
One can see from the figure that the two prescriptions for factorization breaking result in rather distinct shapes of the $x_{\gamma}$ distribution, which 
supports its potential to discriminate between these two scenarios. 

The ATLAS measurement of inclusive dijet photoproduction in Pb-Pb UPCs was performed in the so-called ``0nXn'' event topology,
which required a particular number of forward neutrons in zero degree calorimeters (ZDCs), namely, no neutrons
in one direction and one or more neutrons in the opposite direction. This condition has totally eliminated the contribution of coherent nuclear diffraction,
which is part of nuclear PDFs. It raises the practical question of the magnitude of the diffractive contribution to inclusive dijet photoproduction in UPCs.

\begin{figure}[t]
  \centerline{%
    \includegraphics[width=8cm]{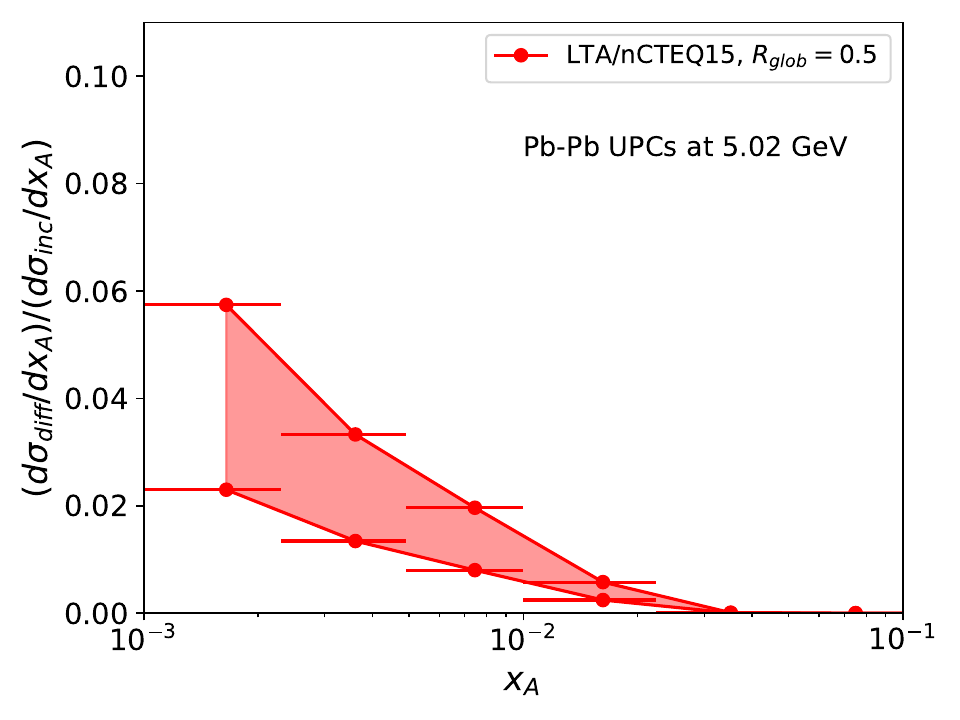}
    \includegraphics[width=8cm]{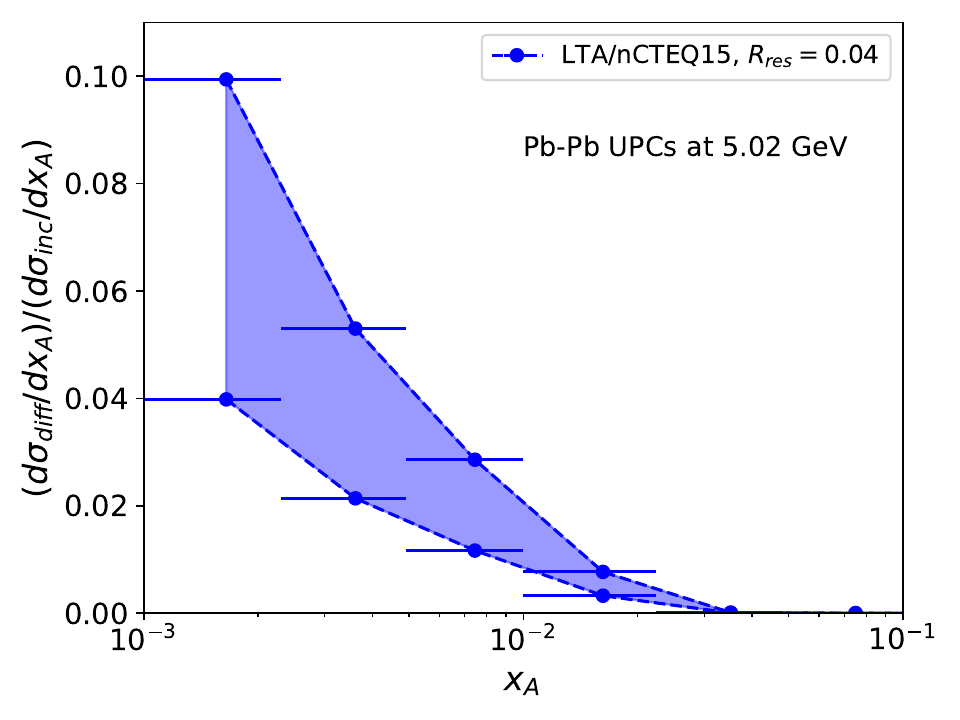}}
  \caption{NLO pQCD predictions for the ratio of the cross sections of diffractive to inclusive dijet photoproduction in Pb-Pb UPCs at 5.02 
  TeV as a function of $x_{A}$. The calculation uses LTA predictions for nuclear diffractive PDFs (the shaded bands represent the theoretical
  uncertainty), the nCTEQ15 nuclear PDFs, and two scenarios
  of the QCD factorization breaking in hard diffraction realized through the global suppression factor $R_{\rm glob}=0.5$ (left panel) and the direct photon suppression factor $R_{\rm res}=0.04$ (right panel).
}
  \label{fig:diff_tot_xA}
\end{figure}

The analysis~\cite{Guzey:2020ehb} has shown that the diffractive contribution to inclusive dijet photoproduction in Pb-Pb UPCs in
the ATLAS kinematics does not exceed $5-10$\% at small $x_A$. It is illustrated in Fig.~\ref{fig:diff_tot_xA} showing the
ratio of the cross sections of diffractive to inclusive dijet photoproduction in Pb-Pb UPCs at 5.02 TeV, $(d\sigma_{\rm diff}/dx_A)/(d\sigma_{\rm inc}/dx_A)$, as a a function of $x_A$. 
The calculation uses the LTA predictions for nuclear diffractive PDFs entering the diffractive cross section
(the shaded bands quantity the theoretical uncertainty, see Fig.~\ref{fig:dsigma_diff}) and the nCTEQ15 nuclear PDFs
for the inclusive cross section. The effect of the QCD factorization breaking for hard diffraction (see the discussion above) is included
through either the global suppression factor $R_{\rm glob}=0.5$ (left panel) or the direct photon suppression factor $R_{\rm res}=0.04$ (right panel).
The two scenarios of the factorization breaking lead to different magnitudes and shapes of the dependence of $(d\sigma_{\rm diff}/dx_A)/(d\sigma_{\rm inc}/dx_A)$ on $x_A$.

The small value of the $(d\sigma_{\rm diff}/dx_A)/(d\sigma_{\rm inc}/dx_A)$ ratio is predominantly an effect of the restricted kinematics with large 
$p_{T,1} > 20$ GeV and not-sufficiently small $x_A > 0.001$ and large relative suppression of nuclear diffractive PDFs by leading twist nuclear shadowing~\cite{Guzey:2024xpa}. Thus, the diffractive contribution and the ensuing ambiguity in the determination of
the photon-emitting nucleus (ambiguity in the invariant photon-nucleon energy $W_{\gamma p}$) can be safely neglected in the kinematics
of the ATLAS measurement.

To enhance the diffractive signal, one needs to expand the kinematic coverage by primarily lowering $p_T$ of jets. For instance, using $p_{T,1} > 10$ GeV and $p_{T,2} > 5-7$ GeV, one can reach
$(d\sigma_{\rm diff}/dx_A)/(d\sigma_{\rm inc}/dx_A))=10-20$\% at $x_A \approx 5 \times 10^{-4}$. In the case of $pp$ UPCs at 13 TeV, where the collision energy is larger and there is no nuclear suppression of diffractive PDFs by nuclear shadowing, the ratio of the diffractive and inclusive cross sections 
of dijet production is sizable, $(d\sigma_{\rm diff}/dx_p)/(d\sigma_{\rm inc}/dx_p) \approx 10-15$\% at $x_p \approx 5 \times 10^{-4}$. 

\section{Summary and outlook}
\label{sec:conclusions}

Photoproduction of jets is a standard tool of perturbative QCD, which provides important information on parton distributions
of the proton, the real photon, and nuclei. Within the framework of collinear factorization of pQCD, NLO calculations describe well 
the available data on dijet photoproduction in $ep$ scattering at HERA. Application of this framework to photon-nucleus scattering
in heavy-ion UPCs at the LHC can be used to obtain complementary constraints on nuclear PDFs, measure for the first time nuclear diffractive PDFs
and shed new light on the mechanism of QCD factorization breaking in hard diffraction.

Using NLO pQCD, we calculate the cross section of inclusive dijet photoproduction in Pb-Pb UPCs at 5.02 TeV at the LHC and demonstrate that
it describes well the preliminary ATLAS data. We show that this cross section probes nuclear PDFs down to $x_A \approx 0.005$ and, when used
in the form of pseudo-data in a Bayesian analysis, can reduce the current small-$x$ uncertainties of the state-of-art nuclear PDFs by approximately a factor of 2.

Considering coherent nuclear scattering and using predictions of the leading twist approach (LTA) to nuclear shadowing for nuclear diffractive PDFs, 
we make NLO pQCD predictions for the cross section of diffractive dijet photoproduction in Pb-Pb UPCs at 5.1 TeV. We show that its distribution
in the photon momentum fraction $x_{\gamma}$ is sensitive to both the effect of nuclear shadowing in nuclear diffractive PDFs and
the mechanism of QCD factorization breaking in hard diffraction. In particular, it allows one to discriminate between the two
scenarios of factorization breaking, where its effect is introducing either through the global suppression factor
$R_{\rm glob}=0.5$ or the resolved photon suppression factor $R_{\rm res}=0.04$.
We also show that due to large leading twist nuclear shadowing and restricted ATLAS kinematics (large jet transverse momentum $p_T$ and not sufficiently small $x_A$), the diffractive contribution to the inclusive cross section of dijet photoproduction does not exceed $5-10$\%. It means that one can 
expect only small corrections of the ATLAS (and other similar) data for the excluded diffractive contribution, which simplifies their
interpretation in terms of usual nuclear PDFs.

Both inclusive and diffractive dijet photoproduction in Pb-Pb UPCs at the LHC can be viewed as precursors of analogous measurement in photon-nucleus
scattering at the planned Electron-Ion Collider (EIC)~\cite{Guzey:2020zza,Guzey:2020gkk}.

\section*{Acknowledgments}

The research of V.G.~was funded by the Academy of Finland project 330448, the Center of Excellence in Quark Matter
of the Academy of Finland (projects 346325 and 346326), and the European Research Council project ERC-2018-ADG-835105
YoctoLHC. The research of M.S.~was supported by the US Department of Energy Office
of Science, Office of Nuclear Physics under Award No. DE- FG02-93ER40771.


\section*{References}

\begin{thebibliography}{99}

%\cite{Salam:2010zt}
\bibitem{Salam:2010zt}
G.~P.~Salam,
%``Elements of QCD for hadron colliders,''
[arXiv:1011.5131 [hep-ph]].
%67 citations counted in INSPIRE as of 18 Mar 2024

%\cite{Sapeta:2015gee}
\bibitem{Sapeta:2015gee}
S.~Sapeta,
%``QCD and Jets at Hadron Colliders,''
Prog. Part. Nucl. Phys. \textbf{89}, 1-55 (2016)
%doi:10.1016/j.ppnp.2016.02.002
[arXiv:1511.09336 [hep-ph]].
%43 citations counted in INSPIRE as of 18 Mar 2024

%\cite{Laenen:2016njs}
\bibitem{Laenen:2016njs}
E.~Laenen,
%``QCD,''
%doi:10.5170/CERN-2016-003.1
[arXiv:1708.00770 [hep-ph]].
%0 citations counted in INSPIRE as of 18 Mar 2024

%\cite{ZEUS:2012pcn}
\bibitem{ZEUS:2012pcn}
H.~Abramowicz \textit{et al.} [ZEUS],
%``Inclusive-jet photoproduction at HERA and determination of alphas,''
Nucl. Phys. B \textbf{864}, 1-37 (2012)
%doi:10.1016/j.nuclphysb.2012.06.006
[arXiv:1205.6153 [hep-ex]].
%66 citations counted in INSPIRE as of 18 Mar 2024

%\cite{H1:2014cbm}
\bibitem{H1:2014cbm}
V.~Andreev \textit{et al.} [H1],
%``Measurement of multijet production in $ep$ collisions at high $Q^2$ and determination of the strong coupling $\alpha _s$,''
Eur. Phys. J. C \textbf{75}, no.2, 65 (2015)
%doi:10.1140/epjc/s10052-014-3223-6
[arXiv:1406.4709 [hep-ex]].
%93 citations counted in INSPIRE as of 18 Mar 2024

%\cite{Klein:2008di}
\bibitem{Klein:2008di}
M.~Klein and R.~Yoshida,
%``Collider Physics at HERA,''
Prog. Part. Nucl. Phys. \textbf{61}, 343-393 (2008)
%doi:10.1016/j.ppnp.2008.05.002
[arXiv:0805.3334 [hep-ex]].
%96 citations counted in INSPIRE as of 18 Mar 2024

%\cite{Newman:2013ada}
\bibitem{Newman:2013ada}
P.~Newman and M.~Wing,
%``The Hadronic Final State at HERA,''
Rev. Mod. Phys. \textbf{86}, no.3, 1037 (2014)
%doi:10.1103/RevModPhys.86.1037
[arXiv:1308.3368 [hep-ex]].
%75 citations counted in INSPIRE as of 18 Mar 2024

%\cite{ZEUS:2005iex}
\bibitem{ZEUS:2005iex}
S.~Chekanov \textit{et al.} [ZEUS],
%``An NLO QCD analysis of inclusive cross-section and jet-production data from the zeus experiment,''
Eur. Phys. J. C \textbf{42}, 1-16 (2005)
%doi:10.1140/epjc/s2005-02293-x
[arXiv:hep-ph/0503274 [hep-ph]].
%245 citations counted in INSPIRE as of 18 Mar 2024

%\cite{NNPDF:2021njg}
\bibitem{NNPDF:2021njg}
R.~D.~Ball \textit{et al.} [NNPDF],
%``The path to proton structure at 1\% accuracy,''
Eur. Phys. J. C \textbf{82}, no.5, 428 (2022)
%doi:10.1140/epjc/s10052-022-10328-7
[arXiv:2109.02653 [hep-ph]].
%285 citations counted in INSPIRE as of 18 Mar 2024

%\cite{Hou:2019efy}
\bibitem{Hou:2019efy}
T.~J.~Hou, J.~Gao, T.~J.~Hobbs, K.~Xie, S.~Dulat, M.~Guzzi, J.~Huston, P.~Nadolsky, J.~Pumplin and C.~Schmidt, \textit{et al.}
%``New CTEQ global analysis of quantum chromodynamics with high-precision data from the LHC,''
Phys. Rev. D \textbf{103}, no.1, 014013 (2021)
%doi:10.1103/PhysRevD.103.014013
[arXiv:1912.10053 [hep-ph]].
%550 citations counted in INSPIRE as of 18 Mar 2024

%\cite{Eskola:2021nhw}
\bibitem{Eskola:2021nhw}
K.~J.~Eskola, P.~Paakkinen, H.~Paukkunen and C.~A.~Salgado,
%``EPPS21: a global QCD analysis of nuclear PDFs,''
Eur. Phys. J. C \textbf{82}, no.5, 413 (2022)
%doi:10.1140/epjc/s10052-022-10359-0
[arXiv:2112.12462 [hep-ph]].
%90 citations counted in INSPIRE as of 18 Mar 2024

%\cite{Slominski:2005bw}
\bibitem{Slominski:2005bw}
W.~Slominski, H.~Abramowicz and A.~Levy,
%``NLO photon parton parametrization using ee and ep data,''
Eur. Phys. J. C \textbf{45}, 633-641 (2006)
%doi:10.1140/epjc/s2005-02458-7
[arXiv:hep-ph/0504003 [hep-ph]].
%25 citations counted in INSPIRE as of 18 Mar 2024

%\cite{Hentschinski:2022xnd}
\bibitem{Hentschinski:2022xnd}
M.~Hentschinski, C.~Royon, M.~A.~Peredo, C.~Baldenegro, A.~Bellora, R.~Boussarie, F.~G.~Celiberto, S.~Cerci, G.~Chachamis and J.~G.~Contreras, \textit{et al.}
%``White Paper on Forward Physics, BFKL, Saturation Physics and Diffraction,''
Acta Phys. Polon. B \textbf{54}, no.3, 3-A2 (2023)
%doi:10.5506/APhysPolB.54.3-A2
[arXiv:2203.08129 [hep-ph]].
%34 citations counted in INSPIRE as of 18 Mar 2024

%\cite{Iancu:2023lel}
\bibitem{Iancu:2023lel}
E.~Iancu, A.~H.~Mueller, D.~N.~Triantafyllopoulos and S.~Y.~Wei,
%``Probing gluon saturation via diffractive jets in ultra-peripheral nucleus-nucleus collisions,''
Eur. Phys. J. C \textbf{83}, no.11, 1078 (2023)
%doi:10.1140/epjc/s10052-023-12165-8
[arXiv:2304.12401 [hep-ph]].
%6 citations counted in INSPIRE as of 20 Mar 2024

%\cite{Boussarie:2021ybe}
\bibitem{Boussarie:2021ybe}
R.~Boussarie, H.~M\"antysaari, F.~Salazar and B.~Schenke,
%``The importance of kinematic twists and genuine saturation effects in dijet production at the Electron-Ion Collider,''
JHEP \textbf{09}, 178 (2021)
%doi:10.1007/JHEP09(2021)178
[arXiv:2106.11301 [hep-ph]].
%42 citations counted in INSPIRE as of 18 Mar 2024

%\cite{Butterworth:2005aq}
\bibitem{Butterworth:2005aq}
J.~M.~Butterworth and M.~Wing,
%``High energy photoproduction,''
Rept. Prog. Phys. \textbf{68}, 2773-2828 (2005)
%doi:10.1088/0034-4885/68/12/R03
[arXiv:hep-ex/0509018 [hep-ex]].
%18 citations counted in INSPIRE as of 18 Mar 2024

%\cite{ATLAS:2017kwa}
\bibitem{ATLAS:2017kwa}
 [ATLAS],
``Photo-nuclear dijet production in ultra-peripheral Pb+Pb collisions,''
ATLAS-CONF-2017-011.
%43 citations counted in INSPIRE as of 18 Jan 2024

%\cite{ATLAS:2022cbd}
\bibitem{ATLAS:2022cbd}
 [ATLAS],
``Photo-nuclear jet production in ultra-peripheral Pb+Pb collisions at $\sqrt{s}_\text{NN} = 5.02$ TeV with the ATLAS detector,''
ATLAS-CONF-2022-021.
%12 citations counted in INSPIRE as of 18 Jan 2024

%\cite{Baltz:2007kq}
\bibitem{Baltz:2007kq}
A.~J.~Baltz, G.~Baur, D.~d'Enterria, L.~Frankfurt, F.~Gelis, V.~Guzey, K.~Hencken, Y.~Kharlov, M.~Klasen and S.~R.~Klein, \textit{et al.}
%``The Physics of Ultraperipheral Collisions at the LHC,''
Phys. Rept. \textbf{458}, 1-171 (2008)
%doi:10.1016/j.physrep.2007.12.001

%\cite{Flett:2020duk}
\bibitem{Flett:2020duk}
C.~A.~Flett, A.~D.~Martin, M.~G.~Ryskin and T.~Teubner,
%``Very low $x$ gluon density determined by LHCb exclusive $J/\psi$ data,''
Phys. Rev. D \textbf{102}, 114021 (2020)
%doi:10.1103/PhysRevD.102.114021
[arXiv:2006.13857 [hep-ph]].
%27 citations counted in INSPIRE as of 20 Mar 2024

%\cite{Guzey:2013xba}
\bibitem{Guzey:2013xba}
V.~Guzey, E.~Kryshen, M.~Strikman and M.~Zhalov,
%``Evidence for nuclear gluon shadowing from the ALICE measurements of PbPb ultraperipheral exclusive $J/{\psi}$ production,''
Phys. Lett. B \textbf{726}, 290-295 (2013)
%doi:10.1016/j.physletb.2013.08.043
[arXiv:1305.1724 [hep-ph]].
%112 citations counted in INSPIRE as of 18 Jan 2024

%\cite{Guzey:2013qza}
\bibitem{Guzey:2013qza}
V.~Guzey and M.~Zhalov,
%``Exclusive $J/{\psi}$ production in ultraperipheral collisions at the LHC: constrains on the gluon distributions in the proton and nuclei,''
JHEP \textbf{10}, 207 (2013)
%doi:10.1007/JHEP10(2013)207
[arXiv:1307.4526 [hep-ph]].
%98 citations counted in INSPIRE as of 18 Jan 2024

%\cite{Frankfurt:2011cs}
\bibitem{Frankfurt:2011cs}
L.~Frankfurt, V.~Guzey and M.~Strikman,
%``Leading Twist Nuclear Shadowing Phenomena in Hard Processes with Nuclei,''
Phys. Rept. \textbf{512}, 255-393 (2012)
%doi:10.1016/j.physrep.2011.12.002
[arXiv:1106.2091 [hep-ph]].
%206 citations counted in INSPIRE as of 18 Jan 2024

%\cite{Strikman:2005yv}
\bibitem{Strikman:2005yv}
M.~Strikman, R.~Vogt and S.~N.~White,
%``Probing small x parton densities in ultraperipheral AA and pA collisions at the LHC,''
Phys. Rev. Lett. \textbf{96}, 082001 (2006)
%doi:10.1103/PhysRevLett.96.082001
[arXiv:hep-ph/0508296 [hep-ph]].
%58 citations counted in INSPIRE as of 17 Jan 2024

%\cite{Klein:2002wm}
\bibitem{Klein:2002wm}
S.~R.~Klein, J.~Nystrand and R.~Vogt,
%``Heavy quark photoproduction in ultraperipheral heavy ion collisions,''
Phys. Rev. C \textbf{66}, 044906 (2002)
%doi:10.1103/PhysRevC.66.044906
[arXiv:hep-ph/0206220 [hep-ph]].
%63 citations counted in INSPIRE as of 14 Feb 2024

%\cite{Goncalves:2009ey}
\bibitem{Goncalves:2009ey}
V.~P.~Goncalves, M.~V.~T.~Machado and A.~R.~Meneses,
%``Heavy Quark Photoproduction in Coherent Interactions at High Energies,''
Phys. Rev. D \textbf{80}, 034021 (2009)
%doi:10.1103/PhysRevD.80.034021
[arXiv:0905.2067 [hep-ph]].
%15 citations counted in INSPIRE as of 26 Feb 2024

%\cite{Goncalves:2017zdx}
\bibitem{Goncalves:2017zdx}
V.~P.~Gon\c{c}alves, G.~Sampaio dos Santos and C.~R.~Sena,
%``Inclusive heavy quark photoproduction in $pp$, $pPb$ and $PbPb$ collisions at Run 2 LHC energies,''
Nucl. Phys. A \textbf{976}, 33-45 (2018)
%doi:10.1016/j.nuclphysa.2018.05.002
[arXiv:1711.04497 [hep-ph]].
%10 citations counted in INSPIRE as of 26 Feb 2024

%\cite{Frixione:1997ks}
\bibitem{Frixione:1997ks}
S.~Frixione and G.~Ridolfi,
%``Jet photoproduction at HERA,''
Nucl. Phys. B \textbf{507}, 315-333 (1997)
%doi:10.1016/S0550-3213(97)00575-0
[arXiv:hep-ph/9707345 [hep-ph]].
%272 citations counted in INSPIRE as of 19 Mar 2024

%\cite{Klasen:1996it}
\bibitem{Klasen:1996it}
M.~Klasen and G.~Kramer,
%``Inclusive two jet production at HERA: Direct and resolved cross-sections in next-to-leading order QCD,''
Z. Phys. C \textbf{76}, 67-74 (1997)
%doi:10.1007/s002880050528
[arXiv:hep-ph/9611450 [hep-ph]].
%101 citations counted in INSPIRE as of 19 Mar 2024

%\cite{Aurenche:2000nc}
\bibitem{Aurenche:2000nc}
P.~Aurenche, L.~Bourhis, M.~Fontannaz and J.~P.~Guillet,
%``NLO Monte Carlo approach in one jet or two jets photoproduction,''
Eur. Phys. J. C \textbf{17}, 413-421 (2000)
%doi:10.1007/s100520000489
[arXiv:hep-ph/0006011 [hep-ph]].
%24 citations counted in INSPIRE as of 19 Mar 2024

%\cite{Klasen:2002xb}
\bibitem{Klasen:2002xb}
M.~Klasen,
%``Theory of hard photoproduction,''
Rev. Mod. Phys. \textbf{74}, 1221-1282 (2002)
%doi:10.1103/RevModPhys.74.1221
[arXiv:hep-ph/0206169 [hep-ph]].
%129 citations counted in INSPIRE as of 19 Mar 2024

%\cite{Guzey:2018dlm}
\bibitem{Guzey:2018dlm}
V.~Guzey and M.~Klasen,
%``Inclusive dijet photoproduction in ultraperipheral heavy ion collisions at the CERN Large Hadron Collider in next-to-leading order QCD,''
Phys. Rev. C \textbf{99}, no.6, 065202 (2019)
%doi:10.1103/PhysRevC.99.065202
[arXiv:1811.10236 [hep-ph]].
%23 citations counted in INSPIRE as of 18 Jan 2024

%\cite{Nystrand:2004vn}
\bibitem{Nystrand:2004vn}
J.~Nystrand,
%``Electromagnetic interactions in nucleus-nucleus and proton-proton collisions,''
Nucl. Phys. A \textbf{752}, 470-479 (2005)
%doi:10.1016/j.nuclphysa.2005.02.051
[arXiv:hep-ph/0412096 [hep-ph]].
%59 citations counted in INSPIRE as of 22 Mar 2024

%\cite{Helenius:2018mhx}
\bibitem{Helenius:2018mhx}
I.~Helenius,
%``Probing nuclear PDFs with dijets in ultra-peripheral Pb+Pb collisions,''
PoS \textbf{HardProbes2018}, 118 (2018)
%doi:10.22323/1.345.0118
[arXiv:1811.10931 [hep-ph]].
%3 citations counted in INSPIRE as of 19 Mar 2024

%\cite{Helenius:2019gbd}
\bibitem{Helenius:2019gbd}
I.~Helenius and C.~O.~Rasmussen,
%``Hard diffraction in photoproduction with Pythia 8,''
Eur. Phys. J. C \textbf{79}, no.5, 413 (2019)
%doi:10.1140/epjc/s10052-019-6914-1
[arXiv:1901.05261 [hep-ph]].
%13 citations counted in INSPIRE as of 19 Mar 2024

%\cite{Kovarik:2015cma}
\bibitem{Kovarik:2015cma}
K.~Kovarik, A.~Kusina, T.~Jezo, D.~B.~Clark, C.~Keppel, F.~Lyonnet, J.~G.~Morfin, F.~I.~Olness, J.~F.~Owens and I.~Schienbein, \textit{et al.}
%``nCTEQ15 - Global analysis of nuclear parton distributions with uncertainties in the CTEQ framework,''
Phys. Rev. D \textbf{93}, no.8, 085037 (2016)
%doi:10.1103/PhysRevD.93.085037
[arXiv:1509.00792 [hep-ph]].
%550 citations counted in INSPIRE as of 19 Mar 2024

%\cite{Gluck:1991jc}
\bibitem{Gluck:1991jc}
M.~Gluck, E.~Reya and A.~Vogt,
%``Photonic parton distributions,''
Phys. Rev. D \textbf{46}, 1973-1979 (1992)
%doi:10.1103/PhysRevD.46.1973
%822 citations counted in INSPIRE as of 19 Mar 2024

%\cite{Guzey:2019kik}
\bibitem{Guzey:2019kik}
V.~Guzey and M.~Klasen,
%``Constraints on nuclear parton distributions from dijet photoproduction at the LHC,''
Eur. Phys. J. C \textbf{79}, no.5, 396 (2019)
%doi:10.1140/epjc/s10052-019-6905-2
[arXiv:1902.05126 [hep-ph]].
%19 citations counted in INSPIRE as of 20 Mar 2024

%\cite{Guzey:2016tek}
\bibitem{Guzey:2016tek}
V.~Guzey and M.~Klasen,
%``Diffractive dijet photoproduction in ultraperipheral collisions at the LHC in next-to-leading order QCD,''
JHEP \textbf{04}, 158 (2016)
%doi:10.1007/JHEP04(2016)158
[arXiv:1603.06055 [hep-ph]].

%\cite{Guzey:2024xpa}
\bibitem{Guzey:2024xpa}
V.~Guzey and M.~Strikman,
%``Suppression of diffraction in deep-inelastic scattering on nuclei and dynamical mechanism of leading twist nuclear shadowing,''
[arXiv:2403.08342 [hep-ph]].
%0 citations counted in INSPIRE as of 21 Mar 2024

%\cite{Collins:1997sr}
\bibitem{Collins:1997sr}
J.~C.~Collins,
%``Proof of factorization for diffractive hard scattering,''
Phys. Rev. D \textbf{57}, 3051-3056 (1998)
[erratum: Phys. Rev. D \textbf{61}, 019902 (2000)]
%doi:10.1103/PhysRevD.61.019902
[arXiv:hep-ph/9709499 [hep-ph]].
%490 citations counted in INSPIRE as of 31 Jan 2024

%\cite{Klasen:2010vk}
\bibitem{Klasen:2010vk}
M.~Klasen and G.~Kramer,
%``Suppression factors in diffractive photoproduction of dijets,''
Eur. Phys. J. C \textbf{70}, 91-106 (2010)
%doi:10.1140/epjc/s10052-010-1436-x
[arXiv:1006.4964 [hep-ph]].
%10 citations counted in INSPIRE as of 22 Mar 2024

%\cite{Guzey:2016awf}
\bibitem{Guzey:2016awf}
V.~Guzey and M.~Klasen,
%``A fresh look at factorization breaking in diffractive photoproduction of dijets at HERA at next-to-leading order QCD,''
Eur. Phys. J. C \textbf{76}, no.8, 467 (2016)
%doi:10.1140/epjc/s10052-016-4304-5
[arXiv:1606.01350 [hep-ph]].
%10 citations counted in INSPIRE as of 21 Mar 2024

%\cite{Guzey:2020ehb}
\bibitem{Guzey:2020ehb}
V.~Guzey and M.~Klasen,
%``How large is the diffractive contribution to inclusive dijet photoproduction in ultraperipheral collisions at the LHC?,''
Phys. Rev. D \textbf{104}, no.11, 114013 (2021)
%doi:10.1103/PhysRevD.104.114013
[arXiv:2012.13277 [hep-ph]].
%7 citations counted in INSPIRE as of 22 Mar 2024

%\cite{Guzey:2020zza}
\bibitem{Guzey:2020zza}
V.~Guzey and M.~Klasen,
%``Next-to-leading order QCD predictions for dijet photoproduction in lepton-nucleus scattering at the future EIC and at possible LHeC, HE-LHeC, and FCC facilities,''
Phys. Rev. C \textbf{102}, no.6, 065201 (2020)
%doi:10.1103/PhysRevC.102.065201
[arXiv:2003.09129 [hep-ph]].
%9 citations counted in INSPIRE as of 22 Mar 2024

%\cite{Guzey:2020gkk}
\bibitem{Guzey:2020gkk}
V.~Guzey and M.~Klasen,
%``Diffractive dijet photoproduction at the EIC,''
JHEP \textbf{05}, 074 (2020)
%doi:10.1007/JHEP05(2020)074
[arXiv:2004.06972 [hep-ph]].
%11 citations counted in INSPIRE as of 22 Mar 2024




\end{thebibliography}
\end{document}